\documentclass[a4paper,11pt]{article}
\usepackage{pos}

\usepackage{hyperref}%
\title{Modelling uncertainties of $t\overline{t}W$ in multi-lepton channel}
\author*[a]{Jasmina Nasufi}
\affiliation[a]{Institute for Theoretical Particle Physics and Cosmology
RWTH-Aachen University, D-52056 Aachen, Germany}

\emailAdd{jasmina.nasufi@rwth-aachen.de}
\abstract{ We compare fixed order and parton shower matched predictions for the process  ${pp\rightarrow \ell^+\nu_{\ell} \ell^-\bar{\nu}_{\ell} \ell^{\pm}\overset{\textbf{\fontsize{2pt}{2pt}\selectfont(---)}}{\nu}_{\ell} b\bar{b}+X}$ at NLO in QCD, including the orders $\mathcal{O}(\alpha_s^3\alpha^6)$ and $\mathcal{O}(\alpha_s\alpha^8)$. The comparison is performed at the integrated and differential fiducial level at the LHC with $\sqrt{s}=13$ TeV. In the absence of parton shower matching procedure that includes the full off-shell effects for this process at NLO in QCD, we propose a new prescription. It enables the inclusion of approximate full off-shell effects to currently available parton shower matched predictions at NLO in QCD.}

\FullConference{%
  The Tenth Annual Conference on Large Hadron Collider Physics - LHCP2022\\
  16-20 May 2022\\
  online
}

\newcommand{\ttX}[1]{$t\overline{t}#1$}
\newcommand{\ttwqcd}{\ttX{W}~QCD}
\newcommand{\ttwew}{\ttX{W}~EW}
\newcommand{\helac}{\textsc{Helac-Nlo}}
\newcommand{\powheg}{\textsc{Powheg-Box}}
\newcommand{\mg}{MG5\_aMC@NLO}
\newcommand{\lodec}{NWA$_{\rm LOdec}$}
\newcommand{\proc}{${pp\rightarrow \ell^+\nu_{\ell} \ell^-\bar{\nu}_{\ell} \ell^{\pm}\overset{\textbf{\fontsize{2pt}{2pt}\selectfont(---)}}{\nu}_{\ell} b\bar{b}+X}$}

\begin{document}
\maketitle

\section{Introduction}
The associated production of a top pair and a $W^{\pm}$ gauge boson is an important SM process. It displays a rich phenomenology as a signal process, with applications such as the charge asymmetry \cite{Maltoni:2014zpa,Frederix:2021zsh,Bevilacqua:2020srb}. Furthermore, it is the dominant background to \ttX{H} in the multi-lepton decay channels \cite{ATLAS:2018mme,CMS:2018uxb,ATLAS-CONF-2019-045,CMS:2020iwy}. Despite good agreement in the SM, a slight tension between experimental measurements and theoretical predictions has been persistently apparent in comparisons so far. This has been attributed to mis-modelling of the process on the theory side. Thus, with the purpose of improving on current theory predictions, we present a direct comparison of state-of-the-art fixed order and parton shower matched predictions in ref. \cite{Bevilacqua:2021tzp}. Additionally we also propose a prescription, which aims to combine the best aspects of the modelling approaches.

\section{Analysis Setup}
We present predictions for ${pp\rightarrow \ell^+\nu_{\ell} \ell^-\bar{\nu}_{\ell} \ell^{\pm}\overset{\textbf{\fontsize{2pt}{2pt}\selectfont(---)}}{\nu}_{\ell} b\bar{b}+X}$ at NLO in QCD. Here $\ell$ labels $\ell\in\{e,\mu\}$ The calculation includes two main contributions, which can be categorized by the coupling order: the NLO QCD correction to the QCD Born at order $\mathcal{O}(\alpha_s^3\alpha^6)$ and the NLO QCD corrections to the pure EW Born  $\mathcal{O}(\alpha_s\alpha^8)$. For ease of notation, they will be labelled \ttwqcd{} and \ttwew{} respectively. We provide fixed order predictions and parton shower matched predictions (NLO+PS). Fixed order predictions include full off-shell predictions, the full NWA and NWA with LO decays. They are generated using the \helac{} software \cite{Cafarella:2007pc,vanHameren:2009dr,Czakon:2009ss,Ossola:2007ax,Bevilacqua:2011xh,Bevilacqua:2013iha,Czakon:2015cla}.  Parton shower matched predictions are generated via \powheg{} \cite{Nason:2004rx,Frixione:2007vw,Frixione:2007zp,Garzelli:2012bn,Honeywell:2018fcl,FebresCordero:2021kcc,Figueroa:2021txg} and \mg{} \cite{Frixione:2002ik,Frixione:2003ei,Artoisenet:2012st,Alwall:2014hca}. Various details are considered to align fixed order and parton shower matched predictions. These are discussed in detail in \cite{Bevilacqua:2021tzp}.\\

\section{Results}

\begin{table}[b]
\begin{center}
\begin{tabular}{l|c|c} 
 $t\bar{t}W^{\pm}$ & \ttwqcd{} & \ttwew{} \\ \hline
 Full off-shell & $1.58^{+3\%}_{-6\%}$ & $0.206^{+22\%}_{-17\%}$ \\
 NWA & $1.57^{+3\%}_{-6\%}$ & $0.190^{+22\%}_{-16\%}$ \\ 
 \lodec{} & $1.66^{+10\%}_{-10\%}$ & $ 0.162^{+22\%}_{-16\%}$ \\ 
 \powheg & $1.40^{+11\%}_{-11\%}$& $0.133^{+21\%}_{-16\%}$\\
 \mg & $1.40^{+11\%}_{-11\%}$&$0.136^{+21\%}_{-6\%}$
\end{tabular}
\caption{Integrated fiducial cross sections for \proc{} at NLO in QCD at order $\mathcal{O}(\alpha_s^3\alpha^6)$ and $\mathcal{O}(\alpha_s\alpha^8)$ ) for various modelling approaches.}
\label{table1}
\end{center}
\end{table}

\begin{figure}[t!]
\includegraphics[width=0.45\linewidth]{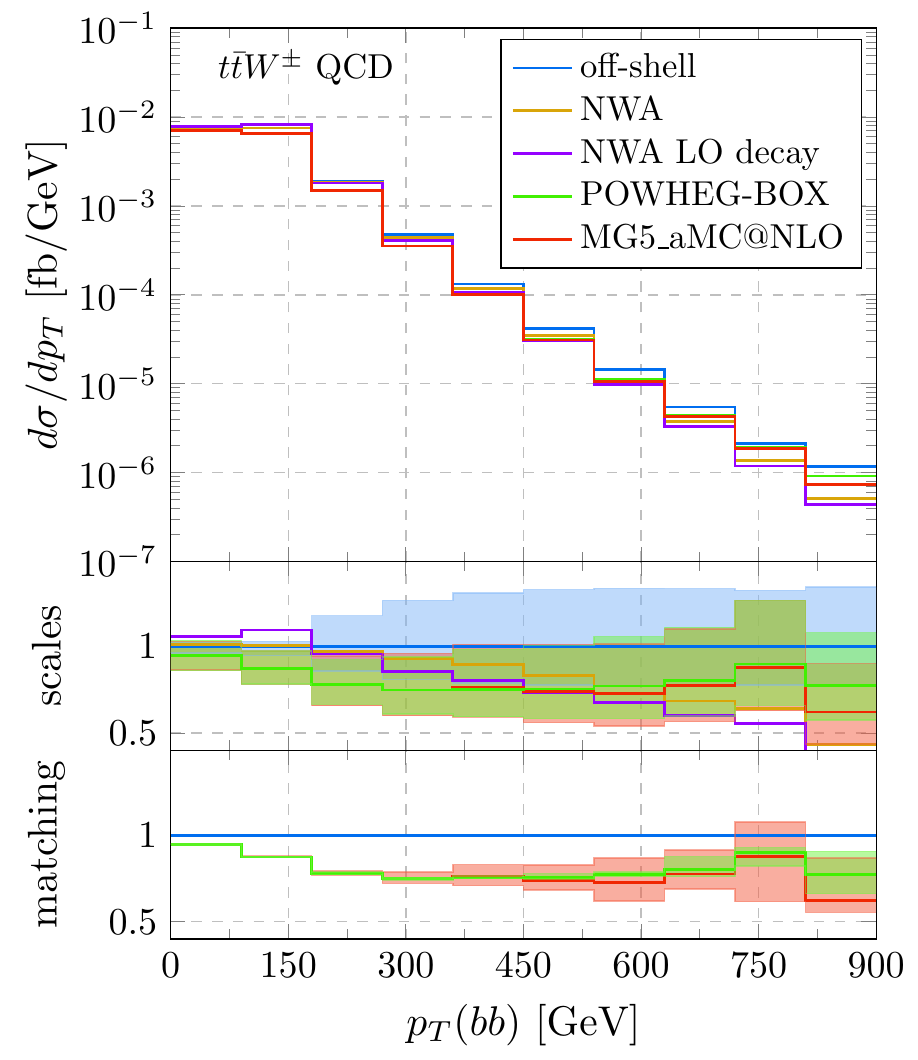}
\includegraphics[width=0.45\linewidth]{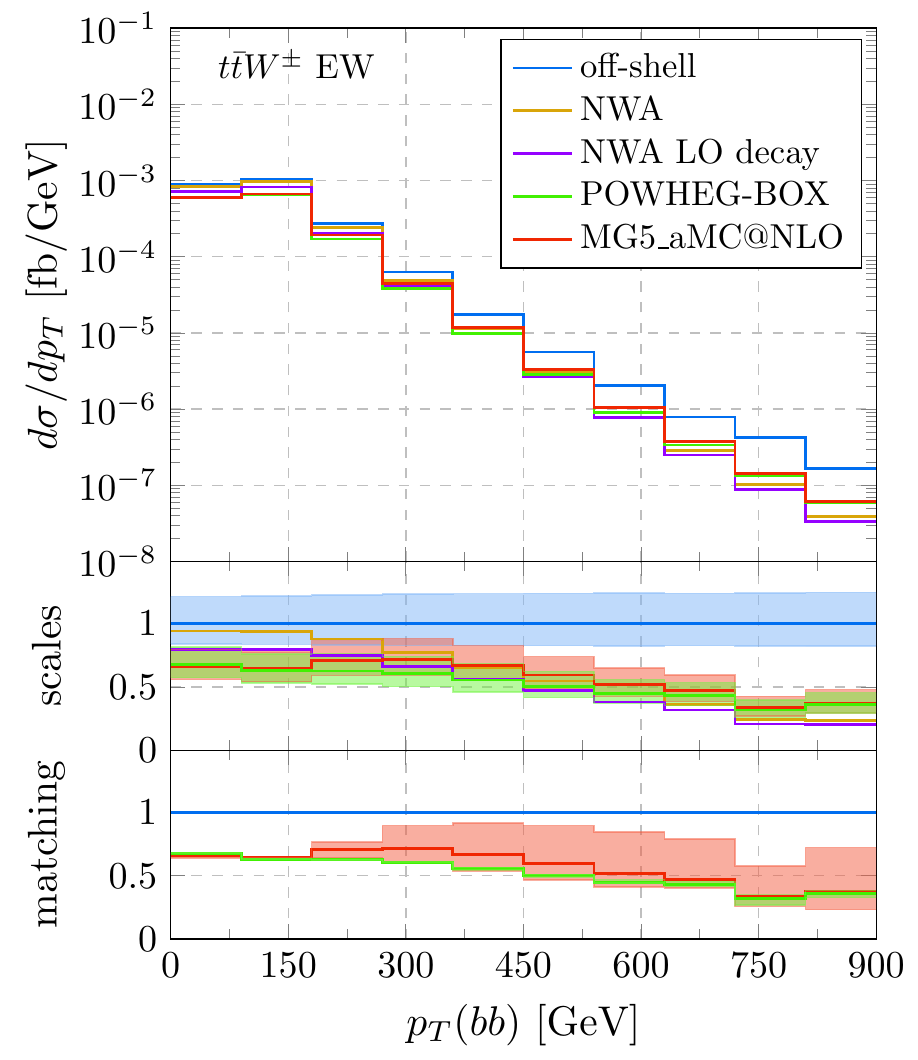}
\caption{Differential fiducial cross section for \proc{} at NLO in QCD at order $\mathcal{O}(\alpha_s^3\alpha^6)$ (left) and $\mathcal{O}(\alpha_s\alpha^8)$ (right). The first panel displays absolute predictions for various modelling approaches, whereas the second panel shows the ratio to the full off-shell predictions with scale uncertainty bands. The last panel displays matching uncertainties for NLO+PS predictions.}
\label{figure1}
\end{figure}

\begin{figure}
\includegraphics[width=0.45\linewidth]{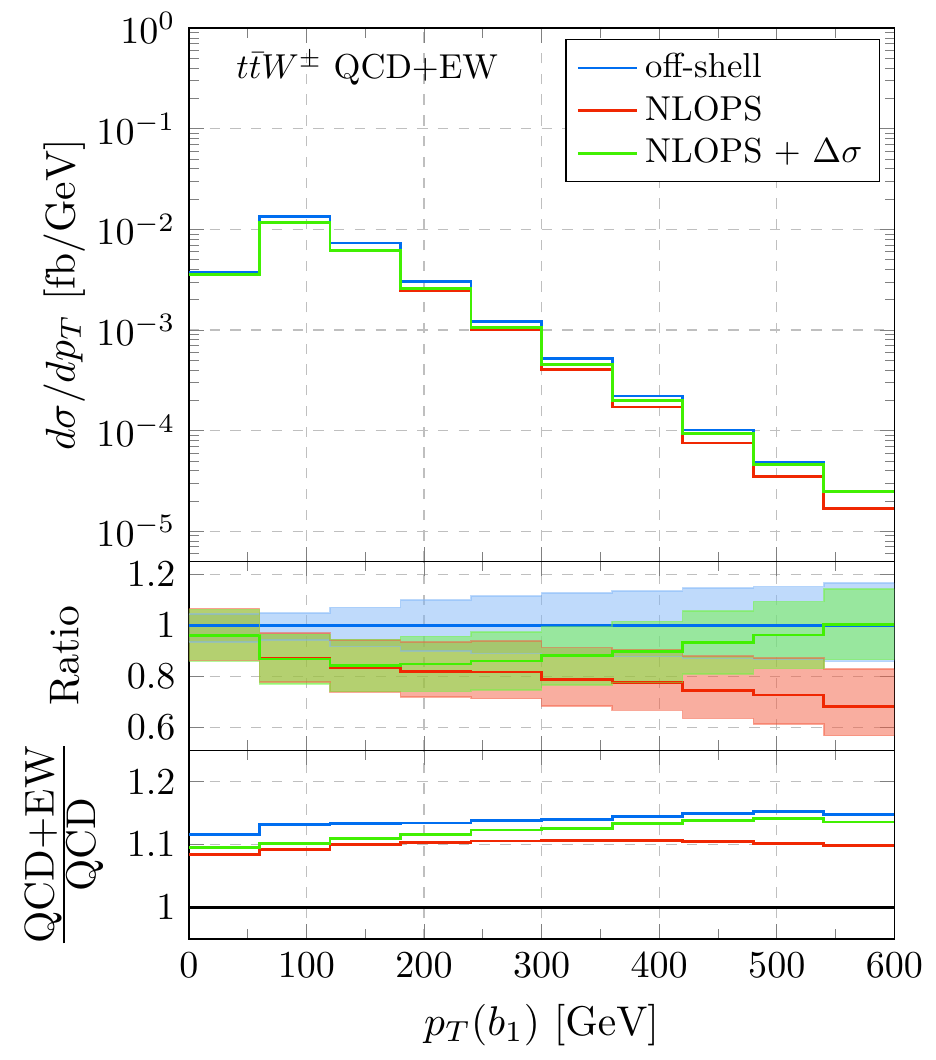}
\includegraphics[width=0.45\linewidth]{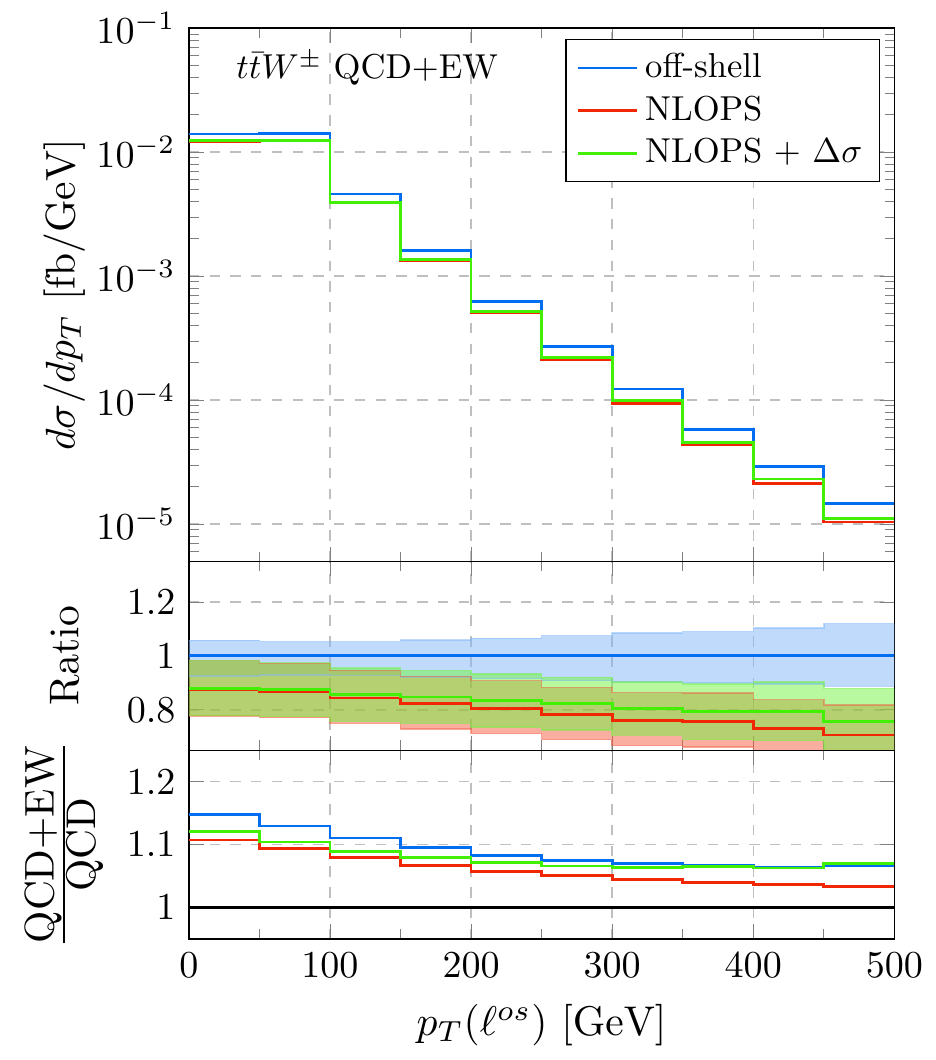}
\caption{Differential fiducial cross section for \proc{} at NLO in QCD including both dominant coupling orders. The first panel shows absolute predictions for various modelling approaches, whereas the second panel shows the ratio to the full off-shell predictions with scale uncertainty bands. The last panel displays the sub-leading \ttwew{} contribution in the ratio.}
\label{figure2}
\end{figure}

Results for the integrated fiducial cross section are shown in table \ref{table1}. The subleading \ttwew{} contribution is sizeable and around $13\%$ of the dominant \ttwqcd{} contribution. The full off-shell effects amount to $0.1\%$ for \ttwqcd{} and a surprising $9\%$ for \ttwew. The enhancement for \ttwew{} is due to $WW\rightarrow WW$ scattering diagrams.  The size of NLO QCD corrections to the decays is $-6\%$ for \ttwqcd{} and $+15\%$ for \ttwew. Neglecting NLO QCD corrections to the decays also impacts the size of the theoretical scale uncertainty for \ttwqcd. It increases from at most $6\%$ for full off-shell predictions and the full NWA, to $10\%$ for \lodec{}. On the other hand, the scale uncertainty for \ttwew{} is unaffected by the modelling and it is LO like, due to the non-trivial $\alpha_s$ dependence appearing at NLO. Parton shower matched predictions generated with \powheg{} and \mg{} are in perfect agreement with each other. They exhibit a similar scale dependence to \lodec{} for both \ttwqcd{} and \ttwew. Compared to fixed order predictions, NLO+PS predictions have a smaller central value because of the multiple radiations.
\begin{table}[b]
\begin{center}
\begin{tabular}{l|c} 
  & $t\bar{t}W^{\pm}$ QCD+EW  \\ \hline
 Full off-shell & $1.79^{+6\%}_{-7\%}$ \\
 NLO+PS & $1.53^{+12\%}_{-11\%}$ \\ 
 NLOPS+$\Delta\sigma$ & $1.56^{+13\%}_{-13\%}$
\end{tabular}
\label{table2}
\caption{Integrated fiducial cross sections for \proc{} at NLO in QCD at order $\mathcal{O}(\alpha_s^3\alpha^6)$ and $\mathcal{O}(\alpha_s\alpha^8)$ ) for various modelling approaches.}
\end{center}
\end{table}
Differential predictions for the transverse momentum of the two hardest $b$-jet system for \ttwqcd{} and \ttwew{} are displayed in figure \ref{figure1}. Full off-shell predictions have a harder high $p_T$ spectrum due to single-resonant contributions, which are not present in any of the other modelling approaches. The discrepancies are more pronounced for \ttwew{}, where the scale uncertainty bands do not overlap. Parton shower matched predictions agree well with each other within scale and matching uncertainties. \\
Parton shower matched predictions can be improved by including full off-shell effects according to:
\begin{eqnarray}
\frac{d\sigma^{\rm th}}{dX} = \frac{d\sigma^{\rm NLO+PS}}{dX} + \frac{d \Delta\sigma_{\rm off-shell}}{dX}, \qquad \frac{d \Delta\sigma_{\rm off-shell}}{dX} = \frac{d\sigma^{\rm NLO}_{\rm off-shell}}{dX} - \frac{d\sigma^{\rm NLO}_{\rm NWA}}{dX}
\label{prescription}
\end{eqnarray}
where $\Delta\sigma_{\rm off-shell}$ is constructed by removing the double resonant contributions from the full off-shell predictions in an approximate way. This prescription has a small impact on the integrated fiducial cross section, which increases by about $2\%$. We expect to see bigger contributions from $\Delta\sigma_{\rm off-shell}$ at the differential level. For this purpose, in figure \ref{figure2} we show the tranverse momentum of the hardest $b$-jet on the left and of the opposite-sign lepton on the right. The improved NLOPS+$\Delta\sigma$ predictions in the bulk of the distributions, whereas towards the high $p_T$ tails they receive full off-shell contributions. These contributions impact $b$-jet observables more than lepton observables.

\section{Conclusions}
In conclusion, in the absence of a resonance aware matching of full off-shell predictions to parton showers at NLO in QCD, we suggest the prescription in eq. \eqref{prescription} for comparisons with unfolded experimental data. 

\begin{center}
\textbf{Acknowledgements}
\end{center}

The work of Jasmina Nasufi was supported by the Deutsche Forschungsgemeinschaft (DFG) under grant 396021762 - TRR 257: P3H - Particle Physics Phenomenology after the Higgs Discovery and under grant 400140256 - GRK 2497: The physics of the heaviest particles
at the Large Hadron Collider.
\bibliography{References} 

\bibliographystyle{JHEP}

\end{document}